\documentclass [12 pt]{article}
\usepackage[psamsfonts]{amssymb}
\usepackage{amsmath}
\author{F. Kheirandish$^{1}$ \footnote{fardin$_{-}$kh@phys.ui.ac.ir} and M.
Amooshahi$^{1}$ \footnote{amooshahi@sci.ui.ac.ir}
\\ $^{1}$ {\small Department of Physics, University of Isfahan,}
\\ {\small Hezar Jarib Ave., Isfahan, Iran.}}
\title{Radiation reaction and quantum damped harmonic oscillator}
\begin{document}
\maketitle
\begin{abstract}
\noindent By taking a Klein-Gordon field as the environment of an
harmonic oscillator and using a new method for dealing with
quantum dissipative systems (minimal coupling method), the quantum
dynamics and radiation reaction for a quantum damped harmonic
oscillator investigated. Applying perturbation method, some
transition probabilities indicating the way energy flows between
oscillator, reservoir and quantum vacuum, obtained.
\end{abstract}
\section{Introduction}
There are some treatments for investigating quantum mechanics
 of dissipative systems, one can consider the
interaction between two systems via an irreversible energy flow
[1,2], or use a phenomenological treatment for a time dependent
Hamiltonian which describes damped oscillations, here we can
refer the interested reader to Caldirola-Kanai Hamiltonian for a
damped harmonic oscillator [3]. There are significant
difficulties about the quantum mechanical solutions of the
Caldirola-Kanai Hamiltonian, for example quantizing with that
Hamiltonian violates the uncertainty relations or canonical
commutation rules, also the
uncertainty relations vanish as time tends to infinity, [4,5,6,7,8]. \\
When a quantum particle of mass $ m $ moving in a one-dimensional
potential $ V(q)$  coupled to a heat bath, the macroscopic
equation describing the time development of the particle motion is
the quantum Langevin equation[9]
\begin{equation}\label{dh.01}
m\ddot{q}+\int_{-\infty}^t
dt'\mu(t-t')\dot{q}(t')+\frac{dV}{dq}=F(t),
\end{equation}
where $ \mu(t)$ is a memory function and $ F(t) $ is a noise
operator associated with absorption of energy of the particle by
heat-bath. This is the Heisenberg equation of motion of the
particle. For a charged particle with charge $ e $,  by taking the
heat-bath as a quantum electromagnetic field, Ford et. al [10],
obtained the Langevin equation (\ref{dh.01}) using the Hamiltonian
\begin{equation}\label{dh.02}
H=\frac{(p-eA_x)^2}{2m}+V(q)+\sum_{\vec{k},s}\hbar c k
a_{\vec{k},s}^\dag a_{\vec{k},s},
\end{equation}
where $ a_{\vec{k},s},a_{\vec{k},s}^\dag $ are annihilation and
creation operators of photon field and $ A_x $ is the $
x$-component of vector potential given by
\begin{equation}
A_x=\sum_{\vec{k},s}(\frac{2\pi\hbar c}{k V})^{\frac{1}{2}}[f^*_k
a_{\vec{k},s}\hat{e}_{\vec{k},s}.\hat{x}+f_k
a_{\vec{k},s}^\dag\hat{e}_{\vec{k},s}.\hat{x}],
\end{equation}
 where $ f_k$ is the form factor, $ \hat{e}_{\vec{k},s}$ are polarization vectors
 and $ V $ is the volume.\\
 In a paper [11] we generalized the Hamiltonian (\ref{dh.02}) for a quantum particle
 which is not necessarily a charged particle and its environment modeled by
a Klein-Gordon type field, the Hamiltonian for a harmonic
oscillator is written as
\begin{equation} \label{d1}
H=\frac{(p-R)^2}{2m}+\frac{1}{2}m\omega^2 q^2+H_B,
\end{equation}
where $m$, $\omega $ are mass and frequency of the oscillator and
$ q $ and $ p $ are position and canonical conjugate momentum
operators of the oscillator, respectively. $ H_B $ is the
Hamiltonian of the reservoir
\begin{equation}\label{d3}
H_B(t)=\int_{-\infty}^{+\infty}d^3k  \omega_{\vec{k}}
b_{\vec{k}}^\dag(t) b_{\vec{k}}(t), \hspace{1.50 cm}
\omega_{\vec{k}}=|\vec{k}|.
\end{equation}
Annihilation and creation operators $ b_{\vec{k}}$,
$b_{\vec{k}}^\dag $, in any instant of time, satisfy the following
commutation relations
\begin{equation}\label{d4}
[b_{\vec{k}}(t),b_{\vec{k}'}^\dag(t)]=\delta(\vec{k}-\vec{k}').
\end{equation}
 Operator $ R $ have the basic role in interaction between oscillator and
reservoir and is defined as
\begin{equation}\label{d5}
R(t)=\int_{-\infty}^{+\infty}d^3k [f(\omega_{\vec{k}})
b_{\vec{k}}(t)+f^*(\omega_{\vec{k}})b_{\vec{k}}^\dag(t)].
\end{equation}
It can be shown easily that combination of Heisenberg equation for
$ q(t) $ and $ b_{\vec{k}} $ lead to
\begin{eqnarray}\label{d9}
&&\ddot{q}+\omega^2q+\int_0^t d
t'\dot{q}(t')\gamma(t-t')=\xi(t),\nonumber\\
&&\gamma(t)=\frac{8\pi}{m}\int_0^\infty d\omega_{\vec{k}}
|f(\omega_{\vec{k}})|^2\omega_{\vec{k}}^3\cos\omega_{\vec{k}} t,\nonumber\\
&&\xi(t)=\frac{i}{m}\int_{-\infty}^{+\infty} d^3 k
\omega_{\vec{k}}(f(\omega_{\vec{k}})b_{\vec{k}}(0)
e^{-i\omega_{\vec{k}}t}-f^*(\omega_{\vec{k}})b_{\vec{k}}^\dag(0)e^{i\omega_{\vec{k}}t}),
\end{eqnarray}
Where we have taken $ \hbar=1 $. It is clear that the expectation
value of $ \xi(t) $ in any eigenstate of $ H_B $, is zero. For the
following special choice of coupling function
\begin{equation}\label{d10}
f(\omega_{\vec{k}})=\sqrt{\frac{\beta}{4\pi^2\omega_{\vec{k}}^3}},
\end{equation}
equation (\ref{d9}) takes the form
\begin{eqnarray}\label{d11}
&&\ddot{q}+\omega^2q+\frac{\beta}{m}\dot{q}=\tilde{\xi}(t),\nonumber\\
&&\tilde{\xi}(t)=i\sqrt{\frac{\beta}{4\pi^2m^2}}\int_{-\infty}^{+\infty}
\frac{d^3 k}{\sqrt{\omega_{\vec{k}}}} (
b_{\vec{k}}(0)e^{-i\omega_{\vec{k}}t}-
b_{\vec{k}}^\dag(0)e^{i\omega_{\vec{k}}t}).
\end{eqnarray}
In QED, a charged particle in quantum vacuum interacts with the
vacuum field and its own field, known as the radiation reaction.
In classical electrodynamics, there is only the radiation reaction
field that acts on a charged particle in the vacuum. The vacuum
and radiation reaction fields have a fluctuation-dissipation
connection and both are required for the consistency of QED. For
example the stability of the ground state, atomic transitions and
lamb shift can only be explained by taking into account both
fields. If self reaction was alone the atomic ground state would
not be stable [12]. When a quantum mechanical system interacts
with the quantum vacuum  field, the coupled Heisenberg equations
for both system and field give us the radiation reaction field,
for example it can be shown that the radiation reaction for a
charged harmonic oscillator is $ \frac{2e^2}{3c^3} $ [12]. In this
paper we investigate the dynamics of a quantum damped harmonic
oscillator interacting with the quantum vacuum and a reservoir by
a minimal coupling method appropriate for dissipative quantum
systems.
\section{Quantum dynamics}
When a damped Harmonic oscillator with charge $ e $, interacts
with the quantum vacuum, the total Hamiltonian can be written
like this
\begin{equation}\label{d37}
H=\frac{(p-R-eA_x)^2}{2m}+\frac{1}{2}m\omega^2q^2+H_B+H_F,
\end{equation}
where $ H_B $ and $ R $, are defined by relations (\ref{d3}) and
(\ref{d5}). $ H_F $ is the vacuum Hamiltonian defined by
\begin{equation}\label{d38}
H_F=\int d^3k\sum_{\lambda=1}^2\omega_{\vec{k}}a_{k\lambda}^\dag
a_{k\lambda},
\end{equation}
 and $ A_x $ is the  $x $-component of the vector potential or the vacuum field which in
 the dipole approximation [15] is
\begin{equation}\label{d39}
A_x=\int_{-\infty}^{+\infty}\frac{d^3 k}{
\sqrt{2(2\pi)^3\omega_{\vec{k}}}}\sum_{\lambda=1}^2\varepsilon_x(
\vec{k},\lambda)[ a_{k\lambda}(t)+a_{k\lambda}^\dag(t)],
\end{equation}
$ a_{k\lambda} $ and $ a_{k\lambda}^\dag $ satisfy the usual
commutation relations
\begin{equation}\label{d39.5}
[a_{k\lambda},a_{k'\lambda'}^\dag]=\delta_{\lambda\lambda'}\delta(\vec{k}-\vec{k'}).
\end{equation}
 Applying Heisenberg equation for operators $ q
$ and $ p $, we find
\begin{equation}\label{d40}
\dot{q}=\frac{p-eA_x-R}{2m},\hspace{2.00 cm} \dot{p}=-m\omega^2q,
\end{equation}
and their combination leads to
\begin{equation}\label{d40.5}
\ddot{q}+\omega^2q=-\frac{\dot{R}}{m}-e\frac{\dot{A_x}}{m},
\end{equation}
 also the Heisenberg equation for $
b_{\vec{k}} $ and $ a_{k\lambda} $ is as follows
\begin{eqnarray}\label{d41}
&&\dot{b}_{\vec{k}}=-i\omega_{\vec{k}}b_{\vec{k}}+i\dot{q}f^*(\omega_{\vec{k}}),\nonumber\\
&&\dot{a}_{k\lambda}=-i\omega_{\vec{k}}a_{k\lambda}+ie\dot{q}
\frac{\varepsilon_x(\vec{k},\lambda)}{\sqrt{2(2\pi)^3\omega_{\vec{k}}}},
\end{eqnarray}
with the following formal solution
\begin{eqnarray}\label{d42}
&&
b_{\vec{k}}(t)=b_{\vec{k}}(0)e^{-i\omega_{\vec{k}}t}+if^*(\omega_{\vec{k}})\int_0^t
d t'e^{-i\omega_{\vec{k}}(t-t')}\dot{q}(t'),\nonumber\\
&&a_{k\lambda}(t)=e^{-i\omega_{\vec{k}}t}a_{k\lambda}
(0)+ie\frac{\varepsilon_x(\vec{k},\lambda)}{\sqrt{2(2\pi)^3\omega_{\vec{k}}}}\int_0^t
d t'e^{-i\omega_{\vec{k}}(t-t')}\dot{q}(t'),
\end{eqnarray}
substituting $ \dot{b}_{\vec{k}} $ and $ \dot{a}_{k\lambda} $ into
the right hand side of (\ref{d40.5}), we obtain
\begin{equation}\label{d43}
\ddot{q}+\omega^2 q+\int_0^t d
t'\gamma(t-t')\dot{q}(t')=\xi(t)-\frac{e}{m}E_0(t)-\frac{e}{m}E_{RR}(t),
\end{equation}
where $ \gamma(t) $ and $ \xi(t) $ are defined by (\ref{d9})
 and $ E_0(t) $ is the vacuum field defined by
\begin{equation}\label{d44}
E_0(t)=i\int_{-\infty}^{+\infty}{d^3
k}\sqrt{\frac{\omega_{\vec{k}}}
{2(2\pi)^3}}\sum_{\lambda=1}^2\varepsilon_x( \vec{k},\lambda)[
a_{k\lambda}(0)^\dag
e^{i\omega_{\vec{k}}t}-a_{k\lambda}(0)e^{-i\omega_{\vec{k}}t}],
\end{equation}
$ E_{RR} $ is the radiation reaction electrical field [15]
\begin{equation}\label{d45}
E_{RR}(t)=\frac{e}{3\pi^2}\int_0^t d
t'\dot{q}(t')\int_0^{+\infty}\omega^2
\cos\omega(t-t')d\omega=-\frac{e}{6\pi}\frac{\partial^3
q}{\partial t^3},
\end{equation}
 now substituting the special choice (\ref{d10}), we find
\begin{equation}\label{d46}
 \ddot{q}+\omega^2
 q+\frac{\beta}{m}\dot{q}+\tau\frac{\partial^3 q}{\partial
 t^3}=E_0(t)+\tilde{\xi}(t),
\end{equation}
where $ \tau=\frac{e^2}{6\pi m} $ and $ \tilde{\xi}(t) $ is
defined by (\ref{d11}).
\section{Transition probabilities}
 Let us write the Hamiltonian (\ref{d37}) as
\begin{eqnarray}\label{d47}
&&H=H_0+H',\nonumber\\
&&H_0=\frac{p^2}{2m}+\frac{1}{2}m\omega^2 q^2+H_F+H_B,\nonumber\\
&&H'=-\frac{R}{m}p-\frac{e}{m}A_x p+\frac{e}{m}A_x
R+\frac{R^2}{2m}+\frac{e^2}{2m}A_x^2.
\end{eqnarray}
Because $ \frac{R^2}{2m} $ is of the second order of damping and
$ \frac{e^2}{2m}A_x $ is small in comparison with $
-\frac{e}{m}A_x p $ , then for sufficiently weak damping, we can
approximate $ H' $ by
\begin{equation}\label{d48}
 H'=-\frac{R}{m}p-\frac{e}{m}A_x p+\frac{e}{m}A_x R,
\end{equation}
and in interaction picture, we can write
\begin{eqnarray}\label{d49}
&&H'_I(t)=e^{iH_0t}H'(0)e^{-iH_0t}=\frac{e}{m}\int_{-\infty}^{+\infty}
d^3 k \int_{-\infty}^{+\infty}\frac{d^3 k'}{
\sqrt{2(2\pi)^3\omega_{\vec{k'}}}}\sum_{\lambda'=1}^2\varepsilon_x(\vec{k'},\lambda')\nonumber\\
&&\times[ f(\omega_{\vec{k}})b_{\vec{k}}(0) a_{k'\lambda'}^\dag(0)
e^{i(\omega_{\vec{k'}}-\omega_{\vec{k}})
t}+f^*(\omega_{\vec{k}})a_{k'\lambda'}(0) b_{\vec{k}}^\dag(0)
e^{i(\omega_{\vec{k}}-\omega_{\vec{k'}})t}]\nonumber\\
&&+i\sqrt{\frac{\omega}{2m}}\int_{-\infty}^{+\infty} d^3 k[
f^*(\omega_{\vec{k}})a b_{\vec{k}}^\dag(0)
e^{i(\omega_{\vec{k}}-\omega) t}-f(\omega_{\vec{k}})a^\dag
b_{\vec{k}}(0) e^{i(\omega-\omega_{\vec{k}})t}]\nonumber\\
&&+ie\sqrt{\frac{\omega}{2m}}\int_{-\infty}^{+\infty}\frac{d^3
k}{\sqrt{2(2\pi)^3\omega_{\vec{k}}}}\sum_{\lambda=1}^2[ a
a_{\vec{k}\lambda}^\dag(0) e^{i(\omega_{\vec{k}}-\omega) t}-a^\dag
a_{\vec{k}\lambda}(0)
e^{i(\omega-\omega_{\vec{k}})t}]\varepsilon_x(\vec{k},\lambda),\nonumber\\
&&
\end{eqnarray}
where we have ignored the terms containing  $
 a a_{k\lambda} ,a b_{\vec{k}} , b_{\vec{k}} a_{k\lambda}$  and their
adjoints according to rotating-wave approximation[13], because in
the first order perturbation these terms at the same time
destroys an excited state of harmonic oscillator, quantum vacuum
and reservoir or create at the same time an excited state of
harmonic oscillator, quantum vacuum and reservoir and so violate
the conservation of energy. The density operator in interaction
picture $ \rho_I $ can be  obtained from [13]
\begin{equation}\label{d27}
\rho_I(t)=U_I(t,t_0)\rho_I(t_0)U_I^\dag(t,t_0),
\end{equation}
where $ U_I $, up to the first order perturbation is
\begin{eqnarray}\label{d27.1}
&&U_I(t,0)=1-i\int_0^t d t_1 H'_I(t_1)=1-\nonumber\\
&&-\sqrt{\frac{\omega}{2m}}\int_{-\infty}^{+\infty}d^3 k
[f(\omega_{\vec{k}}) a^\dag b_{\vec{k}}(0)
e^{\frac{i(\omega-\omega_{\vec{k}}) t}{2}}-f^*(\omega_{\vec{k}})
a b_{\vec{k}}^\dag(0)
e^{\frac{-i(\omega-\omega_{\vec{k}})t}{2}}]\frac{\sin\frac{(\omega-\omega_{\vec{k}})}{2}t}{\frac{(\omega-\omega_{\vec{k}})}{2}}\nonumber\\
&&-e\sqrt{\frac{\omega}{2m}}\int_{-\infty}^{+\infty}\frac{d^3
k}{\sqrt{2(2\pi)^3\omega_{\vec{k}}}}\sum_{\lambda=1}^2\varepsilon_x(\vec{k},\lambda)
[a^\dag a_{k\lambda}(0) e^{\frac{i(\omega-\omega_{\vec{k}})
t}{2}}- a a_{k\lambda}^\dag(0)
e^{\frac{-i(\omega-\omega_{\vec{k}})t}{2}}]\frac{\sin\frac{(\omega-\omega_{\vec{k}})}{2}t}{\frac{(\omega-\omega_{\vec{k}})}{2}}\nonumber\\
&&-i\frac{e}{m}\int_{-\infty}^{+\infty}d^3k\int_{-\infty}^{+\infty}\frac{d^3
k'}{\sqrt{2(2\pi)^3\omega_{\vec{k'}}}}\sum_{\lambda'=1}^2\varepsilon_x(\vec{k'},\lambda')\times\nonumber\\
&&\times[f(\omega_{\vec{k}}) b_{\vec{k}}(0) a_{k'\lambda'}^\dag(0)
e^{\frac{i(\omega_{\vec{k'}}-\omega_{\vec{k}}) t}{2}}+
f^*(\omega_{\vec{k}}) b_{\vec{k}}^\dag(0) a_{k'\lambda'}(0)
e^{\frac{-i(\omega_{\vec{k'}}-\omega_{\vec{k}})t}{2}}]
\frac{\sin\frac{(\omega_{\vec{k'}}-\omega_{\vec{k}})}{2}t}
{\frac{(\omega_{\vec{k'}}-\omega_{\vec{k}})}{2}}.\nonumber\\
\end{eqnarray}
Now let us calculate the transition probability $
|n\rangle_\omega\rightarrow|n-1\rangle_\omega $, which indicates
the way energy flows between the subsystems in some different
cases. The interested reader is referred to N. D. Hari Dass et. al
[14],
 for some related decaying rates obtained trough path integral techniques.\\
 \\
1. If $\rho_I(0)=|n\rangle_{\omega\hspace{0.20 cm}\omega}\langle
n|\otimes|0\rangle_F\hspace{00.20 cm}_F\langle
0|\otimes|0\rangle_{B\hspace{0.20 cm}B} \langle
 0|$, where $ |0\rangle_B $ and $ |0\rangle_F $, are the vacuum state of reservoir and
 quantum vacuum respectively and  $ |n\rangle_\omega $ is an excited state
 of the harmonic oscillator, then substituting $ U_I(t,0) $ from
 (\ref{d27.1}) in (\ref{d27}) and tracing out the degrees of freedom of both reservoir and
 quantum vacuum, we
 obtain the following transition probability for the transition  $
 |n\rangle_\omega\rightarrow|n-1\rangle_\omega $ in long time
 approximation
\begin{eqnarray}\label{d32}
&&\Gamma_{n\rightarrow n-1}=Tr_s[|n-1\rangle_{\omega\hspace{0.20
cm}\omega}\langle n-1| \rho_{sI}(t)]\nonumber\\
&&=\frac{4\pi^2 \omega^3 n t}{m}|f(\omega)|^2+\frac{n\omega^2e^2
t}{6\pi m},
\end{eqnarray}
where $ Tr_s $ denotes tracing over the harmonic oscillator and $
\rho_{sI} $ is defined by tracing out the quantum vacuum and
reservoir $ \rho_{sI}=Tr_{F,B}[\rho_I(t)] $. For special choice
(\ref{d10}) we find
\begin{equation}\label{d33}
\Gamma_{n\rightarrow n-1}=\frac{n\beta t}{m}+\frac{n\omega^2e^2
t}{6\pi m},
\end{equation}
and there is no transition from $ |n\rangle_\omega $ to $
|n-1\rangle_\omega $ in this case. \\ \\
 2. $ \rho_I(0)=|n\rangle_{\omega\hspace{0.20 cm}\omega}\langle
n|\otimes|0\rangle_F\hspace{00.20 cm}_F\langle
0|\otimes|\vec{p}_1,...\vec{p}_j\rangle_{B\hspace{0.20 cm}B}
\langle
 \vec{p}_1,...\vec{p}_j| $ where $ |\vec{p}_1,...\vec{p}_j\rangle_B $ denotes a state of
 reservoir that contains $ j $ quanta with corresponding momenta
 $ \vec{p}_1,...\vec{p}_j $,
 from
 \begin{eqnarray}\label{d33.1}
 &&Tr_B[ b_{\vec{k}}^\dag |\vec{p}_1,...\vec{p}_j\rangle_{B\hspace{0.20 cm}B} \langle
 \vec{p}_1,...\vec{p}_j| b_{\vec{k'}}]=\delta(\vec{k}-\vec{k'}),\nonumber\\
 &&Tr_B[b_{\vec{k}} |\vec{p}_1,...\vec{p}_j\rangle_{B\hspace{0.20 cm}B} \langle
 \vec{p}_1,...\vec{p}_j|b_{\vec{k'}}^\dag ]=
 \sum_{l=1}^j \delta(\vec{k}-\vec{p}_l)\delta(\vec{k'}-\vec{p}_l),
 \end{eqnarray}
 and waiting for a long time, we
 obtain
 \begin{eqnarray}\label{d33.2}
&&\Gamma_{n\rightarrow n-1}=\frac{n\beta t}{m}+\frac{n\omega^2e^2
t}{6\pi m},\nonumber\\
&&\Gamma_{n\rightarrow n+1}=\frac{\beta(n+1) t}{4\pi m \omega^2}
\sum_{l=1}^j \delta( \omega_{\vec{p}_l}-\omega),
\end{eqnarray}
for special choice (\ref{d10}).\\
3. $\rho_I(0)=|n\rangle_{\omega\hspace{0.20 cm}\omega}\langle n|
\otimes |0\rangle_F\hspace{00.20 cm}_F\langle 0|\otimes\rho_B^T $
where  $ \rho_B^T=\frac{e^{\frac{-H_B}{K
T}}}{TR_B(e^{\frac{-H_B}{KT}})} $  is the Maxwell - Boltzman
distribution of reservoir, then using relations
\begin{eqnarray}\label{d34}
&&Tr_B[ b_{\vec{k}}\rho_B^T b_{\vec{k'}}]=Tr_B[ b_{\vec{k}}^\dag
\rho_B^T
b_{\vec{k'}}^\dag]=0,\nonumber\\
&& Tr_b[b_{\vec{k}}\rho_B^T
b_{\vec{k'}}^\dag]=\frac{\delta(\vec{k}-\vec{k'})
}{e^{\frac{\omega_{\vec{k}}}{K T}}-1},\nonumber\\
&&Tr_B[ b_{\vec{k}}^\dag \rho_B^T
b_{\vec{k'}}]=\frac{\delta(\vec{k}-\vec{k'})e^{\frac{\omega_{\vec{k}}}{K
T}}}{e^{\frac{\omega_{\vec{k}}}{K T}}-1},
\end{eqnarray}
one can obtain the transition probabilities
\begin{eqnarray}\label{d36}
&&\Gamma_{n\rightarrow n-1}=\frac{n\beta te^{\frac{\omega}{K
T}}}{m(e^{\frac{\omega}{K T}}-1)}+\frac{n\omega^2e^2
t}{6\pi m},\nonumber\\
&&\Gamma_{n\rightarrow n+1}=\frac{(n+1)\beta
t}{m}\frac{1}{e^{\frac{\omega}{K T}}-1}.
\end{eqnarray}
 4. $ \rho_I(0)=|n\rangle_{\omega\hspace{0.20
cm}\omega}\langle
n|\otimes|\vec{p}_1,\lambda_1,...\vec{p}_j,\lambda_j\rangle_F\hspace{00.20
cm}_F\langle
\vec{p}_1,\lambda_1,...\vec{p}_j,\lambda_j|\otimes|0\rangle_{B\hspace{0.20
cm}B} \langle
 0| $ where $ |\vec{p}_1,\lambda_1...\vec{p}_j,\lambda_j\rangle_B $ denotes a state of
 quantum vacuum that contains $ j $ photon with corresponding momenta
 $ \vec{p}_1,...\vec{p}_j $,
  and polarizations $ \lambda_1,...,\lambda_j $ respectively then by making use of
 \begin{eqnarray}\label{d31}
 &&Tr_F[ a_{k\lambda}^\dag |\vec{p}_1,\lambda_1...\vec{p}_j,\lambda_j\rangle_{F\hspace{0.20 cm}F} \langle
 \vec{p}_1,\lambda_1...\vec{p}_j,\lambda_j| a_{k'\lambda'}]=\delta(\vec{k}-\vec{k'})\delta_{\lambda\lambda'},\nonumber\\
 &&Tr_F[a_{k\lambda} |\vec{p}_1,\lambda_1...\vec{p}_j,\lambda_j\rangle_{F\hspace{0.20 cm}F} \langle
 \vec{p}_1,\lambda_1...\vec{p}_j,\lambda_j|a_{k'\lambda'}^\dag ]=
 \sum_{l=1}^j \delta(\vec{k}-\vec{p}_l)\delta(\vec{k'}-\vec{p}_l)\delta_{\lambda\lambda_l}\delta_{\lambda'\lambda_l}\nonumber\\
 &&
 \end{eqnarray}
 we find
 \begin{eqnarray}\label{d33}
&&\Gamma_{n\rightarrow n-1}=\frac{n\beta t}{m}+\frac{n\omega^2e^2
t}{6\pi m},\nonumber\\
&&\Gamma_{n\rightarrow n+1}=\frac{(n+1)e^2\omega t}{16\pi^2 m}
\sum_{l=1}^j\frac{(
\varepsilon_x(\vec{p}_l,\lambda_l))^2}{\omega_{\vec{p}_l}} \delta(
\omega_{\vec{p}_l}-\omega).
\end{eqnarray}
5. $ \rho_I(0)=|n\rangle_{\omega\hspace{0.20 cm}\omega}\langle
n|\otimes|\vec{p}_1,\lambda_1,...\vec{p}_j,\lambda_j\rangle_F\hspace{00.20
cm}_F\langle
\vec{p}_1,\lambda_1,...\vec{p}_j,\lambda_j|\otimes|\vec{q}_1,...,\vec{q}_j\rangle_{B\hspace{0.20
cm}B} \langle
 \vec{q}_1,...,\vec{q}_j| $  using (\ref{d33.1}) and
 (\ref{d31}) we have
\begin{eqnarray}\label{d33}
&&\Gamma_{n\rightarrow n-1}=\frac{n\beta t}{m}+\frac{n\omega^2e^2
t}{6\pi m},\nonumber\\
&&\Gamma_{n\rightarrow n+1}=\frac{\beta(n+1) t}{4\pi m \omega^2}
\sum_{l=1}^j \delta(
\omega_{\vec{p}_l}-\omega)+\frac{(n+1)e^2\omega t}{16\pi^2 m}
\sum_{l=1}^j\frac{(
\varepsilon_x(\vec{p}_l,\lambda_l))^2}{\omega_{\vec{p}_l}} \delta(
\omega_{\vec{p}_l}-\omega).\nonumber\\
&&
\end{eqnarray}
6. $ \rho_I(0)=|n\rangle_{\omega\hspace{0.20 cm}\omega}\langle
n|\otimes|\vec{p}_1,\lambda_1,...\vec{p}_j,\lambda_j\rangle_F\hspace{00.20
cm}_F\langle \vec{p}_1,\lambda_1,...\vec{p}_j,\lambda_j|\otimes
\rho_B^T $ using (\ref{d34}), gives
\begin{eqnarray}\label{d33}
&&\Gamma_{n\rightarrow n-1}=\frac{n\beta te^{\frac{\omega}{K
T}}}{m(e^{\frac{\omega}{K T}}-1)}+\frac{n\omega^2e^2
t}{6\pi m},\nonumber\\
&&\Gamma_{n\rightarrow n+1}=\frac{(n+1)\beta
t}{m}\frac{1}{e^{\frac{\omega}{K T}}-1}+\frac{(n+1)e^2\omega
t}{16\pi^2 m}
\sum_{l=1}^j\frac{(\varepsilon_x(\vec{p}_l,\lambda_l))^2}{\omega_{\vec{p}_l}}
\delta(\omega_{\vec{p}_l}-\omega).\nonumber\\
&&
\end{eqnarray}
7. $ \rho_I(0)=|n\rangle_{\omega\hspace{0.20 cm}\omega}\langle
n|\otimes \rho_F^T\otimes |0\rangle_B\hspace{00.20 cm}_B\langle0|
$ where $ \rho_F^T $ is the Maxwell-Boltzman distribution of
quantum vacuum, using
\begin{eqnarray}\label{d34.5}
&&Tr_F[ a_{k\lambda}\rho_F^T a_{k'\lambda'}]=Tr_F[
a_{k\lambda}^\dag \rho_F^T
a_{k'\lambda'}^\dag]=0,\nonumber\\
&& Tr_F[a_{k\lambda}\rho_F^T
a_{k'\lambda'}^\dag]=\frac{\delta(\vec{k}-\vec{k'})\delta_{\lambda\lambda'}}
{e^{\frac{\omega_{\vec{k}}}{K T}}-1},\nonumber\\
&&Tr_F[ a_{k\lambda}^\dag \rho_F^T
a_{k'\lambda'}]=\frac{\delta(\vec{k}-\vec{k'})e^{\frac{\omega_{\vec{k}}}{K
T}}}{e^{\frac{\omega_{\vec{k}}}{K T}}-1},
\end{eqnarray}
leads to
\begin{eqnarray}\label{d36.5}
&&\Gamma_{n\rightarrow n-1}=\frac{n\beta t}{m}+\frac{n\omega^2e^2
t}{6\pi m}\frac{e^{\frac{\omega}{K T}}}{e^{\frac{\omega}{K T}}-1},\nonumber\\
&&\Gamma_{n\rightarrow n+1}=\frac{(n+1)e^2\omega^2 t}{6\pi
m}\frac{1}{e^{\frac{\omega}{K T}}-1}.
\end{eqnarray}
8. $ \rho_I(0)=|n\rangle_{\omega\hspace{0.20 cm}\omega}\langle
n|\otimes \rho_F^T\otimes
|\vec{q}_1,...,\vec{q}_j\rangle_B\hspace{00.20
cm}_B\langle\vec{q}_1,...,\vec{q}_j| $  using (\ref{d33.1}), we
find
 \begin{eqnarray}\label{d33.2}
&&\Gamma_{n\rightarrow n-1}=\frac{n\beta t}{m}+\frac{n\omega^2e^2
t}{6\pi m}\frac{e^{\frac{\omega}{K T}}}{e^{\frac{\omega}{K T}}-1},\nonumber\\
&&\Gamma_{n\rightarrow n+1}=\frac{(n+1)e^2\omega^2 t}{6\pi
m}\frac{1}{e^{\frac{\omega}{K T}}-1}+\frac{\beta(n+1) t}{4\pi m
\omega^2} \sum_{l=1}^j \delta( \omega_{\vec{p}_l}-\omega).
\end{eqnarray}
9. Finally for $ \rho_I(0)=|n\rangle_{\omega\hspace{0.20
cm}\omega}\langle n|\otimes \rho_F^T\otimes \rho_B^T $, and using
 (\ref{d34}) and (\ref{d34.5}), we find
\begin{eqnarray}\label{d33.2}
&&\Gamma_{n\rightarrow n-1}=[\frac{n\beta t}{m}+\frac{n\omega^2e^2
t}{6\pi m}]\frac{e^{\frac{\omega}{KT}}}{e^{\frac{\omega}{KT}}-1},\nonumber\\
&&\Gamma_{n\rightarrow n+1}=[\frac{(n+1)e^2\omega^2 t}{6\pi
m}+\frac{(n+1)\beta t}{m}]\frac{1}{e^{\frac{\omega}{K T}}-1}.
\end{eqnarray}
So in all cases the rate of energy transmitted from oscillator to
the environment is a constant.
\subsection{Transition probability between quantum vacuum and reservoir}
Energy flows between reservoir and quantum vacuum with a non zero
probability while the state of oscillator remains unchanged. This
is due to the first integral in (\ref{d49}). For example if $
\rho_I(0)=|n\rangle_\omega \hspace{00.20 cm} _\omega\langle
n|\otimes|\vec{p},r\rangle_F\hspace{00.20 cm}
_F\langle\vec{p},r|\otimes|0\rangle_B\hspace{00.20 cm}_B\langle
0| $, where $ |\vec{p},r\rangle_F $ denotes a state that contains
a photon with momentum $\vec{p}$ and polarization $ r $, then
probability of absorbing $ |\vec{p},r\rangle_F $ by reservoir in
the long time approximation is
\begin{equation}\label{d55}
\frac{e^2\omega_{\vec{p}} t}{2\pi
m^2}|f(\omega_{\vec{p}}|^2(\varepsilon_x(\vec{p},r))^2.
\end{equation}
For the initial state $ \rho_I(0)=|n\rangle_\omega \hspace{00.20
cm} _\omega\langle n|\otimes|0\rangle_F\hspace{00.20 cm}
_F\langle0|\otimes\rho_B^T\ $, the probability that a photon with
momentum $ \vec{p} $ and polarization $ r $  be created, in the
long time approximation, is
\begin{equation}\label{d57}
\frac{e^2\omega_{\vec{p}} t}{2\pi
m^2}|f(\omega_{\vec{p}})|^2(\varepsilon_x(\vec{p},r))^2
\frac{1}{e^{\frac{\omega_{\vec{p}}}{KT}}-1},
\end{equation}
where we have used (\ref{d34}). Also if the initial state is $
\rho_I(0)=|n\rangle_\omega \hspace{00.20 cm} _\omega\langle
n|\otimes|\vec{p},r\rangle_F\hspace{00.20 cm}
_F\langle\vec{p},r|\otimes\rho_B^T $, then the probability that
the mentioned photon be absorbed by the reservoir, is
\begin{equation}\label{d58}
\frac{e^2\omega_{\vec{p}} t}{2\pi
m^2}|f(\omega_{\vec{p}}|^2(\varepsilon_x(\vec{p},r))^2\frac{e^{\frac{\omega_{\vec{p}}}{KT}}}{e^{\frac{\omega_{\vec{p}}}{KT}}-1}.
\end{equation}
If the initial state is $ \rho_I(0)=|n\rangle_\omega \hspace{00.20
cm} _\omega\langle n|\otimes|0\rangle_F\hspace{00.20 cm} _F\langle
0|\otimes|\vec{p}_1,...,\vec{p}_j\rangle_B\hspace{00.20cm }
_B\langle\vec{p}_1,...,\vec{p}_j|  $ , then using (\ref{d33.1}),
it can be shown that the probability of absorbing a photon with
momentum $ \vec{p}$ and polarization $ r $, by the quantum vacuum,
in the long time approximation, is
\begin {equation}\label{d60}
\frac{e^2 t}{8m^2\pi^2
\omega_{\vec{p}}}(\varepsilon_x(\vec{p},r))^2\sum_{l=1}^j|
f(\omega_{\vec{p}_l}|^2\delta(\omega_{\vec{p}_l}-\omega_{\vec{p}}),
\end{equation}
and for $ \rho_I(0)=|n\rangle_\omega \hspace{00.20 cm}
_\omega\langle n|\otimes|\vec{p},r\rangle_F\hspace{00.20 cm}
_F\langle\vec{p},r|\otimes|\vec{p}_1,...,\vec{p}_j\rangle_B\hspace{00.20cm
} _B\langle\vec{p}_1,...,\vec{p}_j|  $, the probability that this
photon be absorbed by the reservoir, is
\begin {equation}\label{d61}
\frac{e^2 t\omega_{\vec{p}}}{2m^2\pi
}(\varepsilon_x(\vec{p},r))^2|f(\omega_{\vec{p}}|^2.
\end{equation}
\section{Concluding remarks}
By taking a Klein-Gordon field as the environment of an harmonic
oscillator and using minimal coupling method, we could obtain a
Heisenberg equation that contained a dissipative term proportional
to velocity and also a radiation reaction electromagnetic field .
In this way we observed that energy can be exchanged between
harmonic oscillator, quantum vacuum and reservoir and the rate of
dissipation of energy of the harmonic oscillator was a constant.

\end{document}